\documentstyle[12pt]{article}
\begin{document}
\begin{center}
{\Large\bf Energy-Information Coupling \\From Classical To Quantum
Physics}\\ \vspace{0.5cm} {Arcangelo Rossi}
\\{\em Dipartimento di Fisica, Universit\`a di Lecce, Via Arnesano, I-73100 Lecce, Italy}
\\{\em e-mail: rossi@le.infn.it}
\end{center}
\begin{abstract}
It is known that there is no possibility of transmitting
information without a certain amount of energy. This is
arbitrarily small in Classical Physics, due to the continuous
nature of the energy parameter, while one cannot reduce that
amount below Planck's energy quanta in Quantum Physics. In short,
one cannot send less than a photon from a place to another when
transmitting a minimum of information. However, as single photons
are never completely defined simultaneously in all their
parameters as position and momentum, their exact contribution to
the information transmitted cannot be known in advance, but only
probabilistically predicted. So the information transmitted is
always blurred before one can - in a way that is only
probabilistically predictable -localize or, alternatively,
determine the momenta of the photons transporting it. In order to
enhance the information content of a message, is it then really
possible to exploit the situation by considering the information
contained in a superposition state before localization or,
alternatively, determination of momenta? It would be so if such
blurred information were richer than a defined one, the former
being vaguer than the latter as it would be compatible with more
possible outputs, and only in this sense more extensive. But if
one defines the information content of a state as the negative
logarithm of the probability of the state itself before
measurement, it is not possible to identify the more {\em a
priori} probable blurred information pertaining to that state with
the bigger one. On the contrary, the more defined information has
the bigger content just because it is the more {\em a priori}
improbable, as only one among different possible alternatives.
Then the idea of exploiting a supposedly enhanced information
content of superposition states in Quantum Computation seems to
contradict directly - but, as we will see, without necessity - the
classical probabilistic definition of information.
\end{abstract}

\vspace{0.2cm}
\section {Introduction: revolutionary ontology}
A now prevailing trend in scientific literature consists in
interpreting quantum concepts and processes in ontological terms,
far beyond the more cautious, instrumentalistic traditional
interpretation maintained, for example, by Niels Bohr
\cite{bohr34}. To the effect of totally overwhelming the classical
physical view by radically transforming all fundamental concepts
of physics in the sense of Thomas S. Kuhn's scientific
revolutions. In fact the trend was already exemplified by Kuhn
himself with reference to modern physics by the transformation of
the classical concept of mass in Relativity Theory \cite{kuhn62}:
in Kuhn's opinion this concept was completely changed by the
transition in the sense that under the same term - mass -
completely different, even uncommeasurable realities were
concealed. Such a radical tranformation would then have taken
place in Quantum Mechanics too, for, according to the now
prevailing opinion, even, for example, the fundamental concept of
probability would have assumed in the new theory a quite new
ontological meaning. Now, in fact, it would imply a true creation
out of nothing of individual properties of quantum objects in a
way that would be totally discontinuous and only predictable in
irreducibly probabilistical terms, which would then be interpreted
in ontological sense as absolute chance, and no longer in purely
epistemic sense, as mere ignorance, as in classical physics
\cite{feynman51}. A true causal discontinuity would take place in
a process of objectivation of properties which would exclude their
objective preexistence. As said, it would not be referred to our
mere ignorance or only partial knowledge of individual properties
as in classical physics, where they were supposed in any case as
pre-existent. This new ontological concept of probability in
Quantum Mechanics has however been largely discussed and it is not
necessary to insist on it here. On the contrary, another
fundamental concept surely deserves more attention for what
concerns the problematicity of its transition from classical to
quantum physics, the concept of information
\cite{shann49}-\cite{brill62}. There is in fact a growing
diffusion of a concept of quantum information, often linked even
to technological applications (to be true, mainly tentative, as
the so called Quantum Computation), thought of as ontologically
distinguished from classical information, to the point of
outlining a true quantum information theory alternative to the
classical one \cite{zeil99}. On purpose, it is mostly fit to
search for which novelties Planck's quantum hypothesis surely
introduced into the concept of information, even though it did not
really modify the classical concept, but, on the contrary,
maintained some of its essential characters. First of all, the
energy-information coupling and, second, the essential link
between probability and information content we will discuss later
on at more length. In this sense, as we will see, the application
of the notion of information content to quantum states does not
imply at all a modification of the classical definition of
information based on that link, but simply invalidates the claim
made by the so called quantum information to gain an information
content richer than what is allowed by the classical information
theory.

\section {The quantum energy-information coupling}
The transmission of information from a broadcasting system to a receiving one,
that locally reduces the entropy of the receiving system, always involves the
transmission of a certain amount of energy, that physically carries out that
information \cite{wiener48}. Neither can information travel as an ordering of
elements, nor can it be picked up as a local entropy reduction factor if at
least a minimum amount of energy is not simultaneously transmitted. Therefore
a piece of information which travels and is picked up is a message carried out
particularly by the energy of light. Sun light is in fact not only energy
poured on our planet but also a carrier of a message that informs us on what
happens in the sun: we are in front of a coupling of both energy transmission
between communicating systems (as, in this case, sun and earth) and
information transmission between the systems themselves. In short, they are
connected not only energetically but also informationally, provided that
energy is the widely more important element, as it is what literally sustains
and supports the message of which information consists, that is the ordering
of elements constituting it.

However, because energy was conceived of as a continuous parameter
in Classical Physics, it was maintained there that an arbitrarily
small amount of energy could carry out information in any case,
provided it met a receiving device able to get information, even
if this was carried out by extremely small, even vanishing amounts
of energy. Energy-information coupling between communicating
systems was then maintained possible even through arbitrarily
small energy exchanges. However, in Quantum Physics energy is
exchanged in finite minimum amounts: line disturbances normally
taking place in telephone communications are then not reducible
with continuity until annihilation, beneath the discrete energy
size of the electrons constituting the current: their destructive
power is therefore strong up to destroying the information carried
out by the current itself. The energy carrying out the signal must
be then at least as strong as counterbalancing the noise energy in
order not to be lost \cite{wiener50}. Even the main information
carrier which is, as mentioned, light, has a quantum structure.
Light of a certain frequency is radiated  in indivisible
curpuscules, light quanta or photons, endowed with an energy
corresponding to that frequency. No energy (and then no
information coupled with it) can be transmitted which is less than
a single photon energy of a certain frequency, corresponding  to a
minimum indivisible energy for that frequency: it is really
unconceivable that  a piece of information can be carried out by
less than a photon for a certain frequency. In any case, that very
small amount of energy is enough to realize an effective
information coupling between communicating systems. Even a photon
can in fact allow, or better carry out a consistent transfer of
information, as in the photoelectric effect. That will however
depend on the physical properties of the photon itself as, in
particular, position and momentum, defining its information
content through reciprocal relations between their values.

\section {Quantum information?}
Here we arrive at the crucial point related to what has to be
changed, if any, in the concept of information through the
transition from classical to quantum physics. In fact, we know by
Quantum Mechanics that, if  we try to exactly define one of these
conjugate complementary properties, the other will result no more
exactly defined, and their reciprocal relation will be then liable
to different possible definitions. There will be not only one but
different possible definitions of the relations between the values
of the conjugate complementary properties according to different
possible actualizations, through following measures, of the
property still undetermined  compared to the previously exactly
defined one \cite{green97}. As a consequence, the information
relating to the superposition state of the possible values of the
second property before measurement, will have a content smaller
than the one referring to the defined state after measurement; for
the former is, as said before, less defined and then more
compatible with different possible results of the measurement
itself than the latter, if one accepts Shannon's \cite{shann49}
and Wiener's \cite{wiener48} classical definition of the
information content of a state as the negative logarithm of the
probability of that state. The quantum superposition state is in
fact no doubt more {\em a priori} probable, as it is compatible
with a larger number of possible measurement results, than the
state following the measurement, which is on the contrary only one
of different {\em a priori} possible states. The sum of squared
moduli of possible values of the property in superposition state
before measurement, taken as the measure of the {\em a priori}
probability of the state itself, is in fact equal to 1
\cite{ghir97} and then its negative logarithm, taken as the
measure of its information content, cannot in any case be more
than 0.

However all that remarkably contrasts with the idea underlying the
so called quantum computation, according to which the information
content of a superposition state is richer than that of a definite
state, so allowing a parallel calculus  which simultaneously
utilizes the ensemble of the possible values of the undefined
observable before measurement. As a consequence, a new information
unit is even defined besides the traditional bit, that is the
qubit, as a unit which corresponds to a plurality of possible
alternative simultaneous choices, instead of a single possible
choice between two alternatives (yes or not)
\cite{deutsch97}-\cite {preskill99}. As underlined by supporters
of this view \cite{bruck99}, it corresponds to interpreting the
information content of quantum measurements as the measure of real
actualizations of even non-preexisting properties starting from a
certain state. It has then no more to do with revelations of
preexisting values (yes/not) of properties as measure of their
{\em a priori} probability in the classical sense.

\section {Quantum computation or classical parallel calculus?}
But if you want to maintain Shannon's and Wiener's classical
definition of the information content of a state (for a different
definition referring it to the  {\em a priori} probability of the
state itself in a different way is not so straightforward),  you
cannot avoid suspecting that the richer information utilized by
the so called Quantum Computation in order to develop a parallel
calculus more powerful than the traditional serial one, is only
due to the simultaneous carrying out of several distinguished
serial computation lines. Their total information content will be
then still measured, according to the classical definition, by the
negative logarithm of the probability of the several distinguished
corresponding states. As this probability is obtained by simply
multiplying one another the probabilities of all these definite
states, that are, as is known, unity fractions, it will be
certainly smaller than the probability of each single state, and
its negative logarithm measuring the total information content of
all states will be then larger than that of each single state
\cite {shann49}.  However, all that seems to have nothing to do
with quantum superposition, particularly with its "fuzzy" or
"unsharp" interpretation, which underlines its absolute novelty
and the unreducibility of its information content to classical
physics. At least, the state ensembles defined by the so called
quantum computation are not demonstrated to be identical  to the
undefined superposition states described by Quantum Mechanics,
apart from the practical consideration that in fact the coherence
of superposition states cannot be maintained for longer than too
short a time in order to implement a durable computation activity
\cite {zurek91}. The last ones are in fact much vaguer,
non-factorizable in as much as they are characterized by
interference terms you cannot find anyway in classical physics, to
the effect that they cannot be reduced to simple products of
composing states, as the "fuzzy" or "unsharp" interpretation
particularly underlines \cite{zadeh65}. Without such
demonstration, the so called quantum parallel calculus seems to be
rather interpretable as a simple product of single serial calculi
simultaneously carried out, which is of course more powerful than
each of them according to Wiener's and Shannon's classical
definition of information, that is then not  necessary to put in
question. In fact, it fully suffices, as said and worthy
repeating, to explain in its probabilistic terms the so called
Quantum Computation as a simultaneous carrying out of several
distinguished serial computation lines, instead of introducing a
mysterious quantum computation based, as seen, on quantum
superposition states, which is in any case poorer, in information
content, than those simultaneous serial computations, according to
Wiener's and Shannon's classical  {\em a priori} probabilistic
definition of information which is still the only viable one.


\begin{thebibliography}{99}

\bibitem{bohr34} N. Bohr, {\em Atomic Theory and the Description of Nature},
Cambridge University Press, Cambridge, London (1934).

\bibitem{kuhn62} T. S. Kuhn, {\em The Structure of Scientific Revolutions},
The University of Chicago Press, Chicago (1962).

\bibitem{feynman51} R. P. Feynman, The concept of probability in quantum
mechanics, in {\em Proceedings of the Second Berkeley Simposium on
Mathematical Statistics and  Probability}, University of
California Press, Berkeley and Los Angeles (1951), 533.

\bibitem{shann49} C. Shannon and W. Weaver, {\em The Mathematical Theory of
Communication}, University of Illinois Press, Urbana (1949).

\bibitem{brill62} L. Brillouin, {\em Science and Information Theory},
Academic Press, New York(1962).

\bibitem{zeil99} A. Zeilinger, A foundation principle for quantum mechanics,
{\em Foundations of Physics}, {\bf 29} (1999), 631.

\bibitem{wiener48} N. Wiener, {\em Cybernetics, or Control and Communication
in the Animal and the Machine}, The Technology Press of M.I.T., Cambridge
Mass. (1948).

\bibitem{wiener50} N. Wiener, {\em The Human Use of Human Beings}, Houghton
Mifflin Company, Boston  (1950).

\bibitem{green97} G. Greenstein and A. G. Zajonk, {\em The Quantum Challenge.
Modern Research on the Foundations of Quantum Mechanics}, Jones
and Bartlett Pbs., Boston (1997).

\bibitem{ghir97} G. Ghirardi, {\em Un'occhiata alle carte di Dio},
il Saggiatore, Milano (1997).

\bibitem{deutsch97} D. Deutsch, {\em The Fabric of Reality}, Penguin Books,
London (1997).

\bibitem{preskill99} J. Preskill, {\em Quantum Information and Computation},
Springer, Berlin (1999).

\bibitem{bruck99} C. Bruckner and A. Zeilinger, Operationally
invariant information in quantum measurements, {\em Physical
Review Letters}, {\bf 83} (1999), 3354.

\bibitem{zurek91}  W. H. Zurek, Decoherence and the transition from quantum
to classical, {\em Physics Today}, {\bf 44} (1991), 36.

\bibitem{zadeh65} L. A. Zadeh, Fuzzy sets, {\em Information and Control}, {\bf 8}
(1965), 338.

\end{thebibliography}
\end{document}